\documentclass[12pt,a4paper]{article}

\usepackage{a4}
\usepackage{amsfonts}
\usepackage{amssymb}
\usepackage{epsfig}
\usepackage{psfig}
\usepackage{psfrag}
\usepackage{dsfont}

\newcommand{\beqn}{\begin{eqnarray}}
\newcommand{\eeqn}{\end{eqnarray}}
\newcommand{\be}{\begin{equation}}
\newcommand{\ee}{\end{equation}}
\newcommand{\non}{\nonumber \\}

\def\st{Stueckelberg~}
\def\stm{StMSSM~}
\def\s1{$s_{\alpha}$}
\def\s2{$s_{\gamma}$}
\def\s3{$s_{\delta}$}
\def\c1{$c_{\alpha}$}
\def\c2{$c_{\gamma}$}
\def\c3{$c_{\delta}$}

\begin{document}
\thispagestyle{empty}

\begin{flushright}
\vspace{-3cm}
{\small 
NUB-TH-3249 \\
MIT-CTP-3504 \\[0cm]
hep-ph/0406167
}
\end{flushright}
\vspace{1cm}

\begin{center}
{\Large\bf  
A Supersymmetric Stueckelberg $U(1)$ Extension \\[.3cm] of the MSSM
}
\end{center}

\vspace{1.5cm}

\begin{center}

{\bf Boris K\"ors}\footnote{e-mail: kors@lns.mit.edu}$^{,*}$
{\bf and Pran Nath}\footnote{e-mail: nath@neu.edu}$^{,\dag}$
\vspace{.5cm}

\hbox{
\parbox{8cm}{
\begin{center}
{\it
$^*$Center for Theoretical Physics \\ 
Laboratory for Nuclear Science \\ 
and Department of Physics \\ 
Massachusetts Institute of Technology \\ 
Cambridge, Massachusetts 02139, USA \\
}
\end{center}
} 
\hspace{-.5cm}
\parbox{8cm}{\begin{center}
{\it
$^\dag$Department of Physics \\ 
Northeastern University \\
Boston, Massachusetts 02115, \\ USA \\
}
\end{center}
}
}
\vspace{.5cm}
\end{center}

\begin{center}
{\bf Abstract} \\
\end{center}
A \st extension of the MSSM with only one abelian 
vector and one chiral superfield as an alternative to an abelian extension 
with Higgs scalars is presented.
The bosonic sector contains a new gauge boson Z$'$  which is a sharp resonance,
and a new CP-even scalar, which combines with the
MSSM Higgs bosons to produce three neutral CP-even massive states. 
The neutral  fermionic sector has two additional fermions which mix with the 
four MSSM neutralinos to produce an extended $6\times 6$ neutralino mass matrix. 
For the case when the LSP is composed mostly of the \st fermions, 
the LSP of the MSSM will be unstable, which leads to
exotic decays of sparticles with many leptons in final states.
Prospects for supersymmetry searches and for dark matter are discussed.

\clearpage
\setcounter{footnote}{0}

%%%%%%%%%%%%%%%%%%%%%%%%%%%%%%%%%%%%%%%%%%%%%%%%%%%%%%%%%%%%%%%%%%%%%%%%%%%%%%%%%%%

The Stueckelberg mechanism \cite{stueck} generates a mass for 
an abelian gauge boson in a gauge invariant, renormalizable way, 
but without utilizing the Higgs mechanism \cite{higgs}. 
Recently, an extension of the Standard Model \cite{gws} was 
proposed \cite{kn}, where the electroweak gauge group $SU(2)_L\times U(1)_Y$
was enlarged by an extra $U(1)_X$ gauge sector, 
whose gauge field couples to an axionic scalar field in the 
way of a \st coupling, giving rise to a massive neutral gauge boson Z$'$. 
This is the simplest realization of the \st 
mechanism in a minimally extended Standard Model. 
The model predicts the presence of a sharp resonance in $e^+e^-$ 
collision, which is distinctly different from other Z$'$ extensions \cite{zpp} 
that appear for instance in grand unified theories or string and 
brane models (but see also \cite{ghiletal}).
In this Letter we further extend this technique and obtain a 
Stueckelberg extension of the minimal supersymmetric Standard Model (StMSSM). 
Since the \st Lagrangian can only accommodate abelian gauge invariance 
it is mostly sufficient to concentrate on the abelian subsector consisting of 
the hypercharge $U(1)_Y$ and the new $U(1)_X$. 
The minimal set of degrees of freedom that has to be added to the MSSM 
consists of the abelian vector multiplet $C$ with components $(C_\mu, \lambda_C,D_C)$ and 
the chiral multiplet $S$ with components $(\chi,\rho+ia,F)$, which we call the \st multiplet. 
Before supersymmetry breaking, 
the two combine into a single massive spin one multiplet, and 
mix with hypercharge and the 3-component of iso-spin, 
as we shall see. Later we will also include a hidden sector whose matter fields may 
carry charge under the $U(1)_X$. 
For the \st Lagrangian we choose
\beqn
{\cal L}_{\rm St} = \int d\theta^2 d\bar{\theta}^2\, \left[ M_1C+M_2B+  S +\bar S \right]^2\ , 
\label{mass}
\eeqn 
where $B$ is the $U(1)_Y$ vector multiplet with components $(B_\mu, \lambda_B, D_B)$. 
The gauge transformations under $U(1)_Y$ and $U(1)_X$ are 
\beqn \label{stgauge}
\delta_Y B = \Lambda_Y + \bar\Lambda_Y\ ,\quad \delta_Y S = - M_2 \Lambda_Y\ , \nonumber\\ 
\delta_X C = \Lambda_X + \bar\Lambda_X\ , \quad \delta_X S = - M_1 \Lambda_X\ . 
\eeqn
The quantities $M_1,\, M_2$ are ``topological'' \cite{AML} 
input parameters of the model. We define $C$ in Wess-Zumino gauge as   
\beqn
C= -\theta\sigma^{\mu}\bar \theta C_{\mu} 
+i\theta\theta \bar\theta \bar \lambda_C  
-i\bar\theta\bar\theta \theta  \lambda_C
+\frac{1}{2} \theta \theta\bar\theta\bar\theta D_C \ , 
\eeqn 
and similarly $B$, while 
$S$ is
\beqn \label{supS}
S
&=&\frac{1}{2}(\rho +ia ) + \theta\chi 
 + i \theta\sigma^{\mu}\bar\theta \frac{1}{2}(\partial_{\mu} \rho 
+i \partial_{\mu} a) 
\\
&&
+~ \theta\theta F +\frac{i}{2} \theta \theta \bar\theta \bar\sigma^{\mu} \partial_{\mu} \chi 
+\frac{1}{8}\theta\theta\bar\theta\bar\theta (\Box \rho+i\Box a) \ . 
\nonumber
\eeqn
Its complex scalar component contains the axionic pseudo-scalar $a$, which is the analogue of the 
real pseudo-scalar that appears in the non-supersymmetric version in \cite{kn}.  
We write ${\cal L}_{\rm St}$ in component 
notation (e.g.\ \cite{kleinetal})
\beqn \label{stueck} 
{\cal L}_{\rm St} &=& - \frac{1}{2}(M_1C_{\mu} +M_2 B_{\mu} +\partial_{\mu} a)^2
- \frac{1}{2} (\partial_\mu \rho)^2 
- i \chi \sigma^{\mu} \partial_{\mu}\bar {\chi} + 2|F|^2 
\nonumber\\
&& 
\hspace{0cm}
+\rho(M_1D_C +M_2 D_B)
+\bar {\chi} (M_1\bar \lambda_C + M_2\bar \lambda_B) 
 + \chi (M_1 \lambda_C + M_2 \lambda_B) \ . 
\label{ls}
\eeqn
For the gauge fields we add the kinetic terms
\beqn 
{\cal L}_{\rm gkin} = 
-\frac{1}{4} ( B_{\mu\nu} B^{\mu\nu} + C_{\mu\nu} C^{\mu\nu} ) 
- i \lambda_B\sigma^{\mu}\partial_{\mu} \bar \lambda_B 
- i \lambda_C \sigma^{\mu}\partial_{\mu} \bar\lambda_C
+\frac{1}{2}( D_B^2 + D_C^2 ) 
\eeqn 
with $C_{\mu\nu}=\partial_\mu C_\nu - \partial_\nu C_\mu$, and similarly for $B$. 
For the matter fields (quarks, leptons, Higgs scalars, plus hidden sector matter) 
chiral superfields  with components $(f_i,z_i,F_i)$ are introduced, defined similar to 
$S$ but $f_i$ with extra factor $\sqrt 2$ and $z_i$ without the extra $\frac12$ of 
Eq.~(\ref{supS}). 
Their Lagrangian is standard, 
\beqn \label{matt}
{\cal L}_{\rm matt}&=& 
%\sum_i \Big[ 
- |D_\mu z_i|^2 - i f_i \sigma^\mu D_\mu \bar f_i 
- \sqrt{2} \left( ig_Y Y z_i \bar f_i \bar \lambda_B 
+ ig_X Q_X z_i \bar f_i \bar \lambda_C 
                   +\, {\rm h.c.}\, \right)
\nonumber\\  
&&
+g_Y D_B (\bar z_i Y z_i) + g_X D_C (\bar z_i Q_X z_i) + |F_i|^2 
\ , 
\eeqn
where $(Y,g_Y)$ and $(Q_X,g_X)$ are the charge operators and coupling constants 
of hypercharge and $U(1)_X$, and 
$D_\mu = \partial_\mu + ig_Y Y B_\mu + ig_X Q_X C_\mu$ the gauge covariant derivative. 
One further has the freedom to add Fayet-Iliopoulos terms 
\beqn 
\xi_B D_B + \xi_C D_C\ , 
\eeqn 
which leads to 
\beqn 
- D_B = \xi_B + M_2\rho + g_Y \sum_i \bar z_i Y z_i \ , \quad 
- D_C = \xi_C + M_1\rho + g_X \sum_i \bar z_i Q_X z_i \ . 
\eeqn 
We find it convenient to replace the two-component Weyl-spinors by 
real four-component Majorana spinors, 
which we label as
$\psi_S = (\chi_\alpha , \bar\chi^{\dot\alpha})^T$, and 
$\lambda_X = (\lambda_{C\alpha},\bar\lambda_C^{\dot\alpha})$ and 
$\lambda_Y = (\lambda_{B\alpha},\bar\lambda_B^{\dot\alpha})$ for the two gauginos, and 
similarly for the matter fields as well. 
Using identities 
\beqn 
\chi \lambda_C + \bar \chi \bar \lambda_C &=& \bar \psi_S \lambda_X \ , \non
\chi \lambda_C - \bar \chi \bar \lambda_C &=& \bar \psi_S \gamma_5 \lambda_X \ , \non
\chi \sigma^\mu \partial_\mu \bar \chi - (\partial_\mu \chi ) \sigma^\mu \bar \chi &=& 
 \bar \psi_S \gamma^\mu \partial_\mu \psi_S  \ , 
\eeqn 
the total Lagrangian of our extension of the MSSM takes the form 
\beqn  \label{delta} 
{\cal L}_{\rm StMSSM} &=& {\cal L}_{\rm MSSM} + \Delta {\cal L}_{\rm St} + \Delta{\cal L}_{\rm hidden} \ , 
\eeqn 
with 
\beqn 
\Delta {\cal L}_{\rm St} &=& -\frac{1}{2}(M_1C_{\mu} +M_2 B_{\mu} +\partial_{\mu} a)^2
-\frac{1}{2} (\partial_\mu \rho)^2
- \frac12 ( M_1^2 + M_2^2) \rho^2 \non
&&
- \frac{i}{2} \bar\psi_S \gamma^{\mu} \partial_{\mu} \psi_S 
- \frac14 C_{\mu\nu}C^{\mu\nu}
- \frac{i}{2} \bar\lambda_X \gamma^\mu \partial_\mu \lambda_X 
+ M_1 \bar \psi_S\lambda_X + M_2\bar\psi_S\lambda_Y \non
&&
- \sum_i \Big[ \left| D_\mu z_i \right|^2 - \left| D_\mu z_i \right|^2_{C_\mu=0} 
+ \rho \Big( g_Y M_2 ( \bar z_i Y z_i ) + g_X M_1 ( \bar z_i Q_X z_i ) \Big) \non
&&
\hspace{1.5cm} 
+ \frac12 g_X C_\mu \bar f_i \gamma^\mu Q_X f_i 
+ \sqrt 2 g_X \Big( iz_i Q_X \bar f_i \lambda_X +\, {\rm h.c.}\, \Big) \Big] \non
&&
- \rho ( M_2 \xi_B + M_1 \xi_C ) - \frac12 \Big[ \xi_C + g_X \sum_i \bar z_i Q_X z_i \Big]^2\ . 
\eeqn
Here and in the following we assume, for simplicity, that all hidden sector fields are 
neutral under the MSSM gauge group, and vice versa, that all fields of the MSSM are 
neutral under the new $U(1)_X$.\footnote{Therefore no coupling 
$g_Y B_\mu \bar f_i \gamma^\mu Y f_i$ appears in (\ref{delta}), and further the 
interaction $g_X C_\mu \bar f_i \gamma^\mu Q_X f_i$ vanishes for the fermions 
of the MSSM, and only involves hidden sector fermions. 
Under these assumptions $\Delta{\cal L}_{\rm hidden}$ will not be 
relevant for our discussion.}
\\

To clarify the particular properties of the \stm let us briefly compare to 
other extensions of the MSSM with an extra $U(1)_X$ gauge field, but which use Higgs fields 
to generate its mass. 
Most of these are immediately distinct, since they involve direct couplings 
between the new gauge field and the MSSM, i.e.\ the new gauge boson is not neutral under 
hypercharge and iso-spin. This imposes much stronger bounds on its mass, usually 
in the range of $1\, {\rm TeV}$ or larger. Such couplings arise for instance in left-right 
symmetric unified models after breaking to the 
electro-weak gauge group, e.g.\ in \cite{Almeida:2004hj,HLM}. 
These models usually involve many more degrees of freedom than the 
minimal \st $U(1)_X$ extension we consider here. 
The Higgs model that would actually come closest to producing the identical effect as the StMSSM, 
a mass for a single extra abelian $U(1)$ gauge boson, 
would consist of adding the $U(1)_X$ to the MSSM plus a Higgs chiral multiplet with 
charges under hypercharge and $U(1)_X$, say both charges $+1$, but otherwise neutral. 
Its action would be given by a single copy of (\ref{matt}), from which one can read that a vacuum 
expectation value $v_H$ for the scalar component of the Higgs multiplet produces 
all the terms in (\ref{stueck}) that involve $M_1$ or $M_2$, with the replacement $M_1 \rightarrow g_X v_H$ and 
$M_2 \rightarrow g_Y v_H$, though the total Lagrangians do not match completely.
This similarity is, however, deceiving, since a single charged chiral multiplet with 
a Lagrangian of the standard form (\ref{matt}) and its implied gauge invariance, would contribute 
to the ABJ gauge anomaly via the usual triangle diagram and spoil anomaly cancellation for the 
hypercharge. Therefore, one is forced to add at least one more Higgs multiplet of opposite 
charge assignments to cancel the anomaly. For the \st multiplet with its 
unusual gauge transformation (\ref{stgauge}) this problem does not arise, since there are no trilinear 
couplings of the form $g_Y B_\mu \chi \sigma^\mu \bar\chi$ in (\ref{stueck}), and 
the \st fermion $\chi$ has zero charge and does not contribute in a triangle diagram. As a conclusion, the minimal 
anomaly-free supersymmetric abelian Higgs model, which would be closest to the StMSSM, differs 
from the latter already at the level of the number of degrees of freedom.
We also note in passing that because of Eq.~(\ref{stgauge}) the chiral superfield $S$ cannot appear 
in the superpotential unlike the usual abelian extensions with Higgs scalars 
(e.g.\ a term $Sh_1h_2$ is not allowed here; compare to \cite{Ellwanger:1997jj}).  
Thus the \stm appears really unique, not only in its theoretical foundation, but also in its 
predictions. \\

In addition to the soft supersymmetry breaking termns of the MSSM we also add a soft mass $\tilde m_X$ for 
the new neutral gaugino $\lambda_X$. In principle, one could also allow 
soft mass terms for $\rho$ and $\psi_S$, but we leave them out as they are not crucial to our discussion. 
Finally, one has to add gauge fixing terms similar to the $R_\xi$ gauge, which remove the 
cross-terms $M_2 B_\mu \partial^\mu a$ etc. together with the usual ones involving 
the Higgs doublets, see \cite{kleinetal,MK}. This completes the Lagrangian of our model. 
As is typically done for the MSSM, we further assume that the FI parameters $\xi_B$ and $\xi_C$ give 
subdominant contributions relative to other sources of supersymmetry breaking. In fact, for the purpose of the present 
analysis, we let $\xi_B$ and $\xi_C$ vanish. We will discuss the modifications due to non-zero FI 
parameters elsewhere. Note that if $\xi_B=\xi_C=0$ at the tree-level, there is no contribution to 
these terms from loop diagrams, since $U(1)_Y$ and $U(1)_X$ are both anomaly-free \cite{Witten:1981nf}. 
\\

We first concentrate on the neutral vector bosons, ordered $(C_\mu,B_\mu,A_\mu^3)$, 
where $A_\mu^3$ is the 3-component of the iso-spin. By giving two out of the three vector bosons masses, 
the Stueckelberg axion $a$ plus one CP-odd component of the Higgs scalars decouple 
after gauge fixing, and disappear from the physical 
spectrum. In order to avoid 
a mass term for the photon it is required that the expectation values 
for all the scalars charged under $U(1)_X$ vanish. 
Thus, we demand all $\langle \bar z_i Q_X z_i \rangle =0$. 
This part of the supersymmetric model is then identical to the non-supersymmetric 
version, that was the subject of \cite{kn}.
After spontaneous electro-weak symmetry breaking the $3\times 3$ neutral vector boson mass matrix 
takes the form
\beqn
%M^2_{ab} = 
\left[
\begin{array}{c|cc}
M_1^2  &  M_1M_2  &  0\\
\hline 
M_1M_2 & M_2^2 + \frac{1}{4} g_Y^2 v^2 & - \frac{1}{4}g_Yg_2 v^2 \\ 
0 & -\frac{1}{4}g_Yg_2 v^2 & \frac{1}{4}g_2^2 v^2
\end{array}
\right]\ .
\eeqn
Here, $v=2M_{\rm W}/g_2=(\sqrt 2 G_F)^{-\frac{1}{2}}$, $M_{\rm W}$ being the mass of the W-boson, 
$G_F$ the Fermi constant. The matrix has a zero eigen value
which corresponds to the photon, and two massive eigen states, 
the Z and Z$'$ bosons.  
As was pointed out in \cite{kn}, it is most convenient to use the two quantities $M=(M_1^2 + M_2^2)^{1/2}$ and 
$\delta = M_2/M_1$ to parametrize the extension of the Standard Model. 
Experimental bounds then only impose $\delta<10^{-2}$, and $M> 150\, {\rm GeV}$, 
which makes the Z$'$ rather light, and a very sharp resonance in $e^+e^-$ annihilation. 
For further details on the gauge boson sector, see \cite{kn}. 
\\ 

We next turn to the scalars of the StMSSM.  
The total scalar potential involves the two Higgs doublets $h_i$ and $\rho$, and 
is given by 
\beqn
{\cal V} &=& \frac{1}{2} (m_1^2-\rho g_Y M_2) |h_1|^2 + \frac{1}{2} (m_2^2 + \rho g_Y M_2) |h_2|^2 
+ (m_3^2 \epsilon_{ij} h_i h_j + {\rm h.c.} ) 
\nonumber\\
&& 
+\frac{1}{2} (M_1^2+M_2^2) \rho^2
+{\cal V}_D\ , 
\eeqn
where ${\cal V}_D$ is the standard MSSM D-term potential that follows immediately 
from (\ref{matt}). Further, for $i=1,2$ we defined  
$m_i^2= m_{h_i}^2+|\mu|^2$, and $m_3^2=|\mu B|$, $\mu$ being the Higgs mixing 
parameter and $B$ the soft bilinear coupling.  
To introduce (real) expectation values for the neutral components of the 
Higgs fields and $\rho$ we replace  
$h_1^0 \rightarrow (v_1+h^0_1)/\sqrt 2,\ h_2^0 \rightarrow (v_2+h^0_2)/\sqrt 2$, with 
$\tan(\beta) = v_2/v_1$, and 
$\rho \rightarrow v_\rho +\rho$, 
with 
\beqn \label{vrho} 
g_Y M_2 v_\rho =  \frac{M_{\rm W}^2 M^2_2}{m_\rho^2} 
\tan^2(\theta_W) \cos(2\beta) 
%- \frac{g_Y M_2}{m_\rho^2} ( \xi_B + \xi_C )  
\ , 
\eeqn
where $m_{\rho}^2 = M_1^2+M_2^2=M^2$, and 
$\theta_W$ is the unmodified weak mixing angle with $\tan(\theta_W)={g_Y}/{g_2}$. 
Due to $|g_Y M_2 v_\rho| < 10^{-4} M_{\rm Z}^2$, 
the modification  
\beqn 
\frac{(g_2^2+g_Y^2)(v_1^2+v_2^2)}{8} = \frac{m_1^2 -m_2^2 \tan^2(\beta)}{\tan^2(\beta)-1} + 
 \frac{g_Y M_2 v_\rho}{\cos(2\beta)} 
\eeqn 
of the electro-weak symmetry breaking constraints is unimportant. 
The \stm does not at all affect the mass of the CP-odd neutral scalar in the MSSM, which is 
\beqn 
m^2_A = -\frac{m_3^2}{\sin(\beta)\cos(\beta)}\ . 
\eeqn 
The three 
CP-even neutral scalars $(h^0_1, h^0_2,\rho)$ mix with mass$^2$ matrix 
($s_{\beta},c_{\beta}=\sin(\beta), \cos(\beta), t_{\theta}=\tan(\theta_W))$
\beqn
\left[
\begin{array}{cc|c} 
M_{0}^2c^2_{\beta} +m_A^2 s^2_{\beta} 
 &  -(M_{0}^2+m_A^2)s_{\beta}c_{\beta}  & -t_{\theta}c_{\beta} M_{\rm W} M_2 \\
 -(M_{0}^2+m_A^2)s_{\beta}c_{\beta}& M_{0}^2s^2_{\beta} +m_A^2 c^2_{\beta}& 
t_{\theta}s_{\beta} M_{\rm W} M_2  \\
\hline
-t_{\theta}c_{\beta} M_{\rm W} M_2  & t_{\theta}s_{\beta} M_{\rm W} M_2 & 
m_{\rho}^2
\end{array}
\right]
\nonumber
\eeqn
where $M_0^2=(g_2^2+g_Y^2)(v_1^2+v_2^2)/4\simeq M_{\rm Z}^2$.  
One can organize the three eigen states as $(H_1^0, H_2^0, \rho_S)$ such that in the 
limit $M_2/M_1 \rightarrow 0$, $(H_1^0, H_2^0, \rho_S) \rightarrow (H^0, h^0, \rho)$, 
where $H^0$ and $h^0$ are the heavy and the light CP-even neutral Higgs of the MSSM.
The new real scalar $\rho_S$ is dominantly $\rho$, but it also carries small components of 
$H^0$ and $h^0$.  
Although the off-diagonal terms proportional to 
$M_{\rm W}M_2$ can be larger than $(100\, {\rm GeV})^2$, the 
corrections to the mass eigen states through mixing are still under control, since the 
ratio $M_{\rm W}M_2/m_\rho^2$ remains small. For a very low mass 
scale $M_1 \sim 10^2\, {\rm GeV}$, $\rho_S$ can be 
directly produced in the $J^{\rm CP}=0^+$ channel. 
Thus there should be three resonances in the $J^{\rm CP}=0^+$ channel
 in $e^+e^-$ collisions in contrast to just two  for the MSSM case.
 The decay of $\rho_S$  into visible MSSM fields will be dominantly into
 $t\bar t$ (or $b\bar b$ 
if $m_{\rho_S}<2m_t$) through the admixture of $H^0$ and $h^0$. 
The partial decay width can be estimated 
%(ignoring interferences, replace $m_t \rightarrow m_b,\ 
%S_{32} \rightarrow S_{31}$ for $b\bar b$) 
%
\beqn
\Gamma(\rho_S\rightarrow t\bar t\, ) &=&
\frac{3m_{\rho_S}}{8\pi}\left[ \frac{m_t S_{32}}{\sqrt 2 M_{\rm W} \sin(\beta)}\right]^2 
\sqrt {1-\frac{4m_t^2}{m^2_{\rho_S}}}\ , \non 
\Gamma(\rho_S\rightarrow b\bar b\, ) &=&
\frac{3m_{\rho_S}}{8\pi}\left[ \frac{m_b S_{31}}{\sqrt 2 M_{\rm W} \sin(\beta)}\right]^2 
\sqrt {1-\frac{4m_b^2}{m^2_{\rho_S}}}\ .
\eeqn
Here $S_{ij}$ are the elements 
of the rotation matrix  that diagonalizes the Higgs mass$^2$ matrix. 
One estimates that $S_{32}$ and $S_{31}$ are  
${\cal O}(M_2/M_1)\sim 0.01$, and thus the $\rho_S$ decay width will 
be in the range of MeV or less, similar to that of Z$'$ \cite{kn}. Such a sharp 
re\-so\-nance can escape detection unless a careful search is carried out.
The total decay width of $\rho_S$ will be broadened, if it can decay into 
hidden sector matter through the much larger coupling $g_X M_1 \rho (\bar z_i Q_X z_i)$ 
in Eq.~(\ref{delta}). The \stm also modifies the D-term contribution to squark
and slepton masses through the term $g_Y M_2 v_\rho ( \bar z_i Y z_i)$ in (\ref{delta}),   
which is typically negligible due to Eq.~(\ref{vrho}). \\ 

Finally, we come to the neutral fermions of the StMSSM. 
Instead of four neutralinos in the MSSM we now have six, 
consisting of the three gauginos, the two Higgsinos $\tilde h_i$, and the extra 
\st fermion $\psi_S$, which we order as $(\psi_S,\lambda_X,\lambda_Y,\lambda_3,\tilde h_1,\tilde h_2)$.
After spontaneous electro-weak symmetry breaking the $6\times 6$
neutralino mass matrix in the above basis is given by 
\beqn
\left[
\begin{array}{cc|cccc}
0 & M_1 & M_2 & 0 & 0 & 0\\
M_1& \tilde m_X & 0 & 0 & 0 & 0 \\ 
\hline
M_2& 0 & \tilde m_1 & 0 & -c_1M_0 & c_2M_0\\
0 & 0 & 0 & \tilde m_2 & c_3M_0 & -c_4M_0 \\
0 & 0 & -c_1M_0  &  c_3M_0 & 0 & -\mu \\
0 & 0 & c_2M_0  &  -c_4M_0 &  -\mu & 0 
\end{array}
\right]\ , 
\eeqn
where $c_1=c_{\beta}s_{\theta}$, $c_2=s_{\beta}s_{\theta}$,
$c_3=c_{\beta}c_{\theta}$, $c_4=s_{\beta}s_{\theta}$.
We label the mass eigen states as
$(\tilde\chi_a^0,\,\tilde\chi_5^0,\,\tilde\chi_6^0),\ a=1,2,3,4$, 
such that in the limit $\delta = M_2/M_1 \rightarrow 0$, 
$\tilde\chi_a^0$ are 
the four eigen states of the MSSM mass matrix in the lower right-hand corner, with  
$m_{\tilde\chi_1^0}<m_{\tilde\chi_2^0}<m_{\tilde\chi_3^0}<m_{\tilde\chi_4^0}$, plus
the eigen states $ \tilde\chi_5^0,\tilde\chi_6^0$ such that 
\beqn
m_{\tilde\chi_5^0},\ m_{\tilde\chi_6^0} ~=~ \sqrt{M_1^{2} +\frac{1}{4}\tilde m_X^{2}}
\pm\frac{1}{2} \tilde m_X + {\cal O}(\delta)
\ , \quad  m_{\tilde\chi_5^0}\ge m_{\tilde\chi_6^0} \  .
\nonumber 
\eeqn
As long as $m_{\tilde\chi_{6}^0} > m_{\tilde\chi_1^0}$, the 
lightest neutralino of the MSSM, $\tilde\chi_1^0$, will still function as the LSP of 
StMSSM. 
However, when 
$m_{\tilde\chi^0_{6}}<m_{\tilde\chi_1^0}$, $\tilde\chi_{\rm St}^0=\tilde\chi_6^0$ 
 becomes the LSP and (with R-parity conservation) a dark matter candidate.
A numerical analysis shows that this can easily be the case for a wide range of parameters.  
This modifies completely the analysis of the decay channels for 
supersymmetric particles into the LSP, since 
the couplings of $\tilde\chi_{\rm St}^0$ to the visible (and to the hidden) 
matter are quite different than those of $\tilde\chi_{1}^0$.
Aside from the issue of dark matter, the supersymmetric signals at 
particle colliders would be drastically modified, and the usual missing energy signals 
no longer apply.
Indeed $\tilde\chi_{1}^0$ would itself be unstable to decay into $\tilde\chi_{\rm St}^0$
by a variety of channels, such as
\beqn
\tilde\chi_1^0~\rightarrow~ {l}_i\bar {l}_i \tilde\chi_{\rm St}^0\ , \quad 
q_j\bar q_j \tilde\chi_{\rm St}^0\ ,\quad 
{\rm Z} \tilde\chi_{\rm St}^0\ , 
\eeqn
where $i(j)$ are lepton (quark) flavors.
The decay lifetime of $\tilde\chi_1^0$ is highly model dependent, involving the 
parameters of the Stueckelberg sector, i.e., $M_2/M_1$, as well as of the MSSM.
An estimate using bounds on $M_2/M_1$ from \cite{kn} gives
$\tau_{\tilde\chi_{1}^0} \sim 10^{-(19\pm 3)}\, {\rm s}$, which implies that 
$\tilde\chi_{1}^0$ will decay in the detection chamber
if $m_{\tilde\chi_6^0} <m_{\tilde\chi_1^0}$. In this circumstance the signatures
for the detection of supersymmetry change drastically as  discussed below. \\ 
 
Because the direct coupling between $\tilde\chi_{\rm St}^0$
and visible matter is weaker by $M_2/M_1$ than  of the
 MSSM neutralinos $\tilde\chi_a^0$, 
sfermions $\tilde f_j$  will first decay dominantly into the 
MSSM neutralinos, i.e., $\tilde f_j\rightarrow  f_j+\tilde\chi_{a}^0$.
This is then followed by the decay of the $\tilde\chi_{a}^0$ with the chain
ending with $\tilde\chi_{\rm St}^0$ in the end product. Typically this
will lead to multi particle  final states often containing many leptons.
For example, the lightest slepton decay can result in a trileptonic final state 
\beqn
\tilde {l}^{-} ~\rightarrow~ l^-+ \tilde\chi_{1}^0 ~\rightarrow~ 
{l^-} + \left\{ {  l_i^-l_i^+ + \{\tilde\chi_{\rm St}^0 \} \atop 
                  q_j\bar q_j + \{\tilde\chi_{\rm St}^0 \} } \right.  \ , 
\nonumber 
\eeqn
where $\{\tilde\chi_{\rm St}^0 \}$ is the missing energy.
A similar situation arises in the decay of the light 
charginos $\tilde\chi_1^{\pm}$ into charged leptons
\beqn
\tilde\chi_1^{-}~\rightarrow~ l^- + \tilde\chi_1^0 +\bar \nu  
~\rightarrow~ l^- + \left\{ { l_i^-l_i^+ + \{\tilde\chi_{\rm St}^0 + \bar\nu \} 
                    \atop q_j\bar q_j + \{\tilde\chi_{\rm St}^0 + \bar\nu \} } 
\right.  \ . 
\nonumber 
\eeqn
As another example, the decay of a squark or a 
gluino will necessarily allow charged leptons in the final states, 
\beqn
\tilde q&\rightarrow& q l\bar l+\{\tilde\chi_{\rm St}^0 \}\ , \non  
\tilde g&\rightarrow& q\bar q l\bar l+\{\tilde\chi_{\rm St}^0 \}\ . 
\eeqn
Using the above chain of decays, the
processes $p+\bar p\rightarrow \tilde\chi_1^0+\tilde\chi_1^0+X,
\tilde\chi_1^{\pm}+\tilde\chi_1^0+X, \tilde\chi_1^{\pm}+\tilde\chi_2^0+X,
\tilde\chi_1^{\pm}+\tilde\chi_1^{\mp}+X$ at the Tevatron collider 
would  lead to multi particle  final states often with many leptons, and similar 
phenomena will occur at the LHC. Thus the conventional signal for supersymmetry
in supergravity unified models where the decay of an off-shell W-boson leads to 
a trileptonic signal \cite{trilep}, 
${\rm W}^{*-}\rightarrow \tilde\chi_1^-\tilde\chi_2^0 \rightarrow l^- l_i\bar l_i + {\rm missing\  energy}$,   
is replaced by a purely leptonic final state, which has seven leptons
and missing energy.
The decay branching ratios of these are model dependent and we leave 
a more complete investigation to a future work \cite{kn2}. \\ 

However, the preceding analysis is already sufficient to demonstrate,
that in the specific scenario considered, where 
$ m_{\tilde\chi_6^0}<m_{\tilde\chi_1^0}$, the supersymmetric 
signals at the Tevatron and at the LHC are drastically altered.  
 A similar situation will hold
for the supersymmetric signals at a linear collider.
Here, the process 
$e^+e^- \rightarrow \tilde\chi_1^{\pm} + \tilde\chi_1^{\mp}$  would
lead to a purely leptonic final state with six leptons and missing energy
which would provide signatures for this kind of a \stm scenario.
Since the nature of physics beyond the standard
model is largely unknown, it is imperative that one considers all
viable scenarios, including the one discussed here, in the exploration
of new physics.
\\

%%%%%%%%%%%%%%%%%%%%%%%%%%%%%%%%%%%%%%%%%%%%%%%%%%%%%%%%%%%%%%%%%%%%%%%%%%%%%%%%%%%%%%%%%%

The work of B.~K.~was supported by the German Science Foundation (DFG) and in part by
funds provided by the U.S. Department of Energy (D.O.E.) under cooperative research agreement
$\#$DF-FC02-94ER40818. The work of P.~N.~was supported in part by
the U.S. National Science Foundation under the grant NSF-PHY-0139967. 

%%%%%%%%%%%%%%%%%%%%%%%%%%%%%%%%%%%%%%%%%%%%%%%%%%%%%%%%%%%%%%%%%%%%%%%%%%%%%%%%%%%%%%%%%%

\end{document}